# The Lorentz Force on Ions in Membrane Channels of Neurons as a Mechanism for Transcranial Static Magnetic Stimulation

Manuel J. Freire, Senior Member, IEEE, and Joaquín Bernal-Méndez, Senior Member, IEEE

*Abstract— Goal:* Transcranial static magnetic stimulation is a novel noninvasive method of reduction of the cortical excitability in certain neurological diseases that, unlike ordinary transcranial magnetic stimulation, makes use of static magnetic fields generated by permanent magnets. The physical principle underlying transcranial magnetic stimulation is well known, that is, the Faraday´s law. By contrast, the physical mechanism that explains the interaction between neurons and static magnetic fields in transcranial static magnetic stimulation remains unclear, which makes it difficult to improve and fine tune the treatment. In the present work it is discussed the possibility that this mechanism might be the Lorentz force exerted on the ions flowing along the membrane channels of neurons. *Methods:* To support this hypothesis, a dimensional analysis it is carried out to compare the Larmor radius of the ions in the presence of a static magnetic field with the dimensions of the cross section of human axons and membrane channels in neurons. *Results:* This analysis shows that whereas a moderate static magnetic field is not expected to affect the ion flux through axons, nevertheless it can affect the ion flux along membrane channels. *Conclusion:* The overall effect of the static magnetic field would be to introduce an additional friction between the ions and the walls of the membrane channels, thus reducing its conductance. *Significance:* Calculations performed by using a Hodgkin-Huxley model demonstrate that even a slight reduction of the conductance of the membrane channels can lead to the suppression of the action potential, thus inhibiting neuronal activity.

## I. Introduction

TRANSCRANIAL magnetic stimulation (TMS) is a well-established noninvasive method of brain stimulation for diagnosis and treatment of neurological diseases that is based on the application of strong and short pulses of magnetic field (typically 1T of amplitude and $300\mu s$ of duration) generated by current-fed coils [1]. The physics underlying TMS is well known and it is based on the induction of currents in neurons by virtue of the Faraday´s law. Protocols for TMS therapy are well established, being the theta-burst protocol the most extended to induce long-lasting neural changes [2]. Transcranial static magnetic stimulation (tSMS) is a novel noninvasive form of brain stimulation, that makes use of static magnetic fields (SMFs) created by permanent magnets to reduce cortical excitability in humans [3][4][5][6]. Experimental evidences show that SMFs of moderate values (tens to hundreds of mT) can interfere with physiological brain functions [3][4][5][6]. There is also experimental evidence of effect produced by even greater SMFs in Magnetic Field Resonance (MRI) exams [7]. Moreover, the interaction of moderate SMFs with excitable membranes of different biological systems has been extensively reported [8][9][10][11][12]. Despite these evidences, a physical mechanism providing a clear explanation for the interaction of moderate SMFs with neurons has not been identified yet. A better understanding of the physic phenomena underlying this interaction would help to increase the efficiency of the tSMS. At a fundamental level, two kinds of physical mechanisms seem to be feasible candidates to provide this explanation: the magnetic behavior of the constituent molecules of excitable membranes in the presence of a SMF, and the interaction between a SMF and moving ions in neurons through the Lorentz force. Within the first perspective, it has been suggested that the reorientation of diamagnetic anisotropic molecules in the cell membrane can be responsible for the influence of moderate SMF on the cell membrane [8][9]. The second hypothesis has been used to investigate, from a theoretical point of view, the influence of SMFs on the ion current that flows along the axon and is associated with the propagation of the action potential (AP) in nerves [13] [14]. From the analysis carried out in [13] [14], it follows that the Lorentz force exerted by moderate SMFs on the ions flowing along nerves cannot appreciably affect the propagation of the AP. Nevertheless, the AP is associated not only with the ion flux along axons but also with the ion flux along membrane channels. Regarding this, it is interesting to note that it has been suggested that ion channels of neurons can be modelled as FET transistors [15]. Also, it is well known that SMFs can affect the performance of FET transistors in MRI preamplifiers due to the Lorentz force in charge carriers [16]. Thus, in the present work, it is discussed the possibility that the AP can be affected by moderate SMF through the Lorentz force exerted on the ions flowing along the membrane channels in neurons. To support this hypothesis, a dimensional analysis is carried out to estimate the ratio between the Larmor radius of the ions in the presence of a SMF with a value typical of the tSMS [3], and the dimensions of the cross section of human axons and membrane channels. Based on this analysis, it is suggested that, although moderate SMFs cannot affect the ion flux through axons, it may affect the ion flux along membrane channels. It is also suggested that the effect of the Lorentz force is to introduce an additional friction between the ions and the walls of the membrane channels. Since the conventional friction between the ions and the walls accounts for almost 2/3 of the conductance value of the channels [17], we conclude that the ultimate effect of the Lorentz force is to reduce significantly the conductance of channels. Results for the AP obtained with a Hodgkin-Huxley (HH) model [18] reveal that a slight reduction of the conductance of the Na channel can lead to the suppression of the AP.

Section II presents an analysis that rules out the effect of Lorentz force associated with moderate SMFs on ions flowing along axons as a cause of neuron inhibition. Also, the ratio between Larmor radius and the diameter of the region of conduction is presented as a suitable benchmark to determine whether Lorentz force can alter the flow of ions. This criterion is employed in section III to show that membrane channels might see its conductance decreased by a Lorentz force such as that created by a moderate SMF, and that the expected decrease can actually suppress the AP. Finally, conclusions are presented in section IV.

## II. ANALYSIS

As it is well known, the Lorentz force is the force exerted on a charged particle moving in the presence of a SMF. Because this force is perpendicular to both the velocity of the particle and the direction of the SMF, it makes the particle to describe a circular trajectory in a plane perpendicular to the SMF. The radius of this trajectory is referred to as the cyclotron radius or Larmor radius, $R_L$, and it is given by $R_L = mv/qB$, where $m$, $v$ and $q$ are the mass, velocity and charge of the particle, respectively, and $B$ is the amplitude of the SMF.

The AP propagating through the axon of neurons is associated with a longitudinal ion current flowing along the axon. In the presence of a SMF, due to the Lorentz force the ions flowing along the axon experience a deflection of their trajectory which produces a transverse current. In [13] it is theoretically analyzed for the first time the order of magnitude of the SMF necessary to produce an appreciable deflection in the longitudinal current associated with the propagation of the AP in the axons of human neurons. The calculations in [13] show that a magnetic field on the order of 25T is necessary to produce a deflection or reduction of 10% in the ion current along the axon. Such a field is several orders of magnitude greater than moderate SMF and even an order of magnitude greater than typical SMF in MRI systems. Moreover, in [14] a deeper analysis estimates the effect of this deflection in the AP by means of a HH model where a term that accounts for the transverse current that appears as a consequence of the deflection is added in the differential equations, this term being proportional to the value of the SMF. In [14] it is defined a ratio $\alpha$ between the transverse current and the longitudinal current, and it is expressed as a relation between the value of the SMF, $B$, and the transverse mobility of the ions, $\mu$, as $B = \alpha/\mu$. The calculations in [14] show that, in particular, a moderate value for the SMF of $B$=11 mT will produce a reduction of 5% (corresponding to $\alpha$=0.05 in [14]) in the longitudinal current in the axon. In [14] it is shown that taking this into account in the HH model, this will cause a suppression of the AP. This result entirely disagrees with the conclusion in [13]. This apparent paradox can be solved by noting that the analysis carried out in [14] assumes an ion mobility of 5 m²/Vs, which is three orders of magnitude larger than values experimentally reported [13]. For example, in [13] the peak axial electric field during the passage of the AP is reported to be $E$=8 V/m and the ion velocity $v_d$=3.3 × 10$^{-2}$ m/s. Therefore the ion mobility is $\mu = v_d/E = 0.004125$ m²/Vs. Assuming this much more realistic value for $\mu$, the required SMF for a reduction of 5% in the longitudinal current in [14] will be 14.7 T, which is closer to the order of magnitude estimated in [13] (i.e., 25 T).

From the above discussion it can be concluded that moderate SMFs cannot affect the propagation of the AP in human axons. This same conclusion can be also drawn from a simple alternative analysis based on the comparison of the Larmor radius with the diameter of the axons. Consider, for example, a sodium (Na) ion, whose mass and charge are: $m = 3.8 \times 10^{-26}$ kg and $q = 1.67 \times 10^{-19}$ C. To estimate the Larmor radius we can assume an ion velocity in the axon of $v_d = 3.3 \times 10^{-2}$ m/s, (i.e., the same value as in [13]) and a SMF of value $B = 164$ mT. This is the value measured by the authors for the same magnet used in tSMS in [3], at a distance of 2 cm from the surface of the magnet, which is the distance between the scalp and the motor cortex. With those assumptions, the Larmor radius is $R_L = mv/qB = 478$Å. This is two orders of magnitude smaller than the typical diameter of the human axon which is 1$\mu m$. Therefore, in the presence of a moderate SMF of 164 mT the ionic current is expected to flow without significant deflection through the axon. Summing up, it can be concluded that due to the different orders of magnitude of the cross section of the axon and the Larmor radius for moderate SMFs, moderate SMFs cannot affect the transport of ions through the axon, in accordance with [13].

The discussion presented above suggests that the comparison between the size of the cross section of the axon and the Larmor radius can be considered as a benchmark to ascertain whether the Lorentz force associated with a given value of SMF affects the ion nerve conduction. In fact, we have just shown that this criterion allows to rule out Lorentz force due to a moderate SMF as the cause of the AP supression in axons. In view of this, in this work we propose an alternative explanation for the effect on the AP of a moderate SMF based on the effect of Lorentz force on the conductance of membrane channels. To underpin this hypothesis, we will use the benchmark index described above to determine whether a moderate SMF can affect the ion flux along membrane channels. In this regard, a key point to be taken into account is that the size of the cross section of ion channels of excitable membranes is several orders of magnitude smaller than the diameter of the axon.

## III. Results and Discussion

In this section a dimensional analysis is carried out to compare the size of the potassium ($K^+$) channel with the Larmor radius of $K^+$ ions for moderate SMFs. To this end, an estimation of the drift velocity of the ions through the channel is required as a first step. Regarding this point, it is important first to determine whether the flow of the ions through the channel can be considered an ohmic process (or ions should be considered ballistic charges instead). Scientific evidence points out that friction caused by the pore shape and wall tortuosity play an important role in the conductance [19] [20]. Therefore, it is reasonable to consider the flow of ions through the channel as an ohmic process. Under this assumption, the amplitude of the current can be written as $I = J \cdot S$, where $S$ is the average cross section of the channel, and the current density $J$ can be written as $J = qnv_d$, where $n$ is the number of ions per unit volume and $v_d$ the drift velocity of ions. Moreover, $n$ can be written as $n = N/V$, where $N$ is the number of ions that can occupy simultaneously the channel and $V$ is the volume of the channel, that can in turn be expressed as $V = SL$, where L is the length of the channel. Therefore, the drift velocity can be expressed as:

$$v_d = \frac{IL}{Nq}. \qquad (1)$$

The $K^+$ channel extends 45Å, with a wide segment of length 23Å and a narrower selectivity filter of radius 1.5 Å and length 12Å where the ions would have to shed its hydrating waters to enter [17] [21]. The selectivity filter contains two $K^+$ ions [19] [21], that is, the number of ions that can occupy simultaneously the selectivity filter is N=2. Since the amplitude of the current is of the order of picoamperes [17], assuming $I = 1$pA and $L$=12Å, $v_d$ can be estimated from (1) as $v_d = 3.75 \times 10^{-3}$ m/s. From this estimation of the drift velocity, and taking into account that the mass of $K^+$ ion is 39.0983 uma = $6.49 \times 10^{-26}$ kg, the corresponding Larmor radius for a SMF of value $B$=164mT can be calculated as: $R_L = mv_d/qB = 93$Å. This value is of the same order of magnitude as the length of the channel, and what it is more important, it is not negligible in comparison with the width of the channel. Therefore, the component of the SMF perpendicular to the axis of the channel will give rise to a Lorentz force acting on the ions which will curve the trajectory of the ions inside the narrow channel. This situation is sketched in Fig. 1.

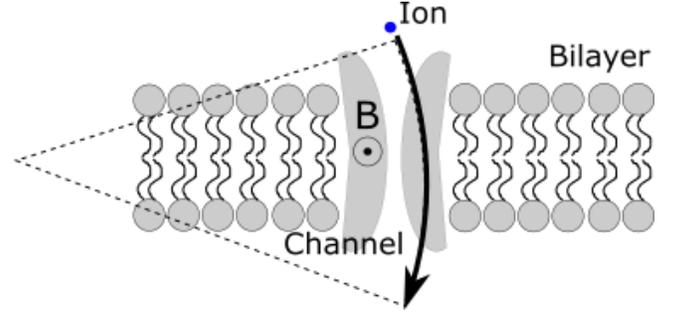

Fig. 1: Sketch of membrane channel and the deflected trajectory of an ion. The Larmor radius is approximately twice the lenght of the channel.

Inside the narrow channels the ions are forced to follow a narrow and straight path. Therefore, the Lorentz force acts pushing the ions against the walls of the channel, which imposes a friction with the walls of the channel. This results in a decrease of the conductance of the ions through the channel.

To estimate to what extent the effect described above can actually decrease the conductivity of the channel it is interesting to revise the relationship between friction, diffusion and conductance. In the Brownian movement, the Einstein relation relates the friction force with the diffusion coefficient $D$ as $D = KT/m\gamma$, $K$ and $T$ being the Boltzmann´s constant and temperature, respectively, and $m\gamma v$ being the friction force in the Langevin's equation [22]. In [17] the diffusion coefficient of $K^+$ in the selectivity filter of the membrane channels is calculated and it is on average 1/3 of the bulk value, whereas in the wider segment of the channel is nearly the same as the bulk value. In the same sense, in [20] it is also reported that the friction is responsible for the diffusion coefficient of $K^+$ to be 3 or 5 times lower than in bulk water ($D = 0.46 \times 10^{-9}$ m²/s in the channel and $2.2 \times 10^{-9}$m²/s in bulk water region). Moreover, in [19] it is pointed out that the different conductance of $K^+$ channels might have different causes, the friction among them. Thus, in [17] it is shown that the reduction of the diffusion coefficient in the selectivity filter (the narrower part of the channel) influences the overall channel conductance. Those evidences suggest that the friction introduced by the Lorentz force in the dynamics of ions through membrane channels can result in the reduction of the conductance of the channels. The analysis was carried out for the $K^+$ channel but the conclusion can be generalized to the rest of channels.

Although the expected reduction of conductance caused by friction due to Lorentz Force is only a factor of 2 or 3, as mentioned above, this reduction might be enough to completely suppress the AP. This is due the fact that the AP generation is quite sensitive to small variations of the conductance values. To illustrate this, Fig. 2 shows changes undergone by the transmembrane potential of a neuronal cell segment in response to three consecutive equal stimuli for three different values of the conductance of the fast Na channel, which is greatly involved in the onset of the AP. These results have been calculated by solving the differential equations of the HH model of AP generation by means of the HHSim software [18], a free graphical simulator that provides access to the parameters of the HH model. Fig. 2 shows three spikes generated under stimuli for three different values of the conductance of the fast Na channel. The first spike corresponds to a conductance of 120 $\mu S$, the second spike corresponds to 80 $\mu S$ and the last spike is for a conductance of 60 $\mu S$. For this last value, it can be observed that, even though the change in conductance is only a 25% with respect to the previous value, the AP is almost entirely suppressed.

## IV. CONCLUSION

In this work it is demonstrated that whereas Lorentz force produced by moderate SMF is not expected to produce appreciable effects on the ions flowing along the axon of neurons, it might well affect the flux of the ions along the membrane channels in neurons. This is due to the different ratios of the cross sections of axons and membrane channels with respect to the corresponding Larmor radius. It has been shown that in the membrane channels the Lorentz force can effectively produce a friction of the ions with the walls of the channel, and that this additional friction might reduce the conductance of the channels. Calculations of neuron responses by using a Hodgkin-Huxley (HH) model have illustrated that reductions of conductance of the same order as those expected can effectively suppress the AP in neurons. The evidences provided by this analysis make of the Lorentz force a feasible candidate to be the main physical mechanism explaining the reduction of the excitability of the motor cortex achieved by the tSMS technique.

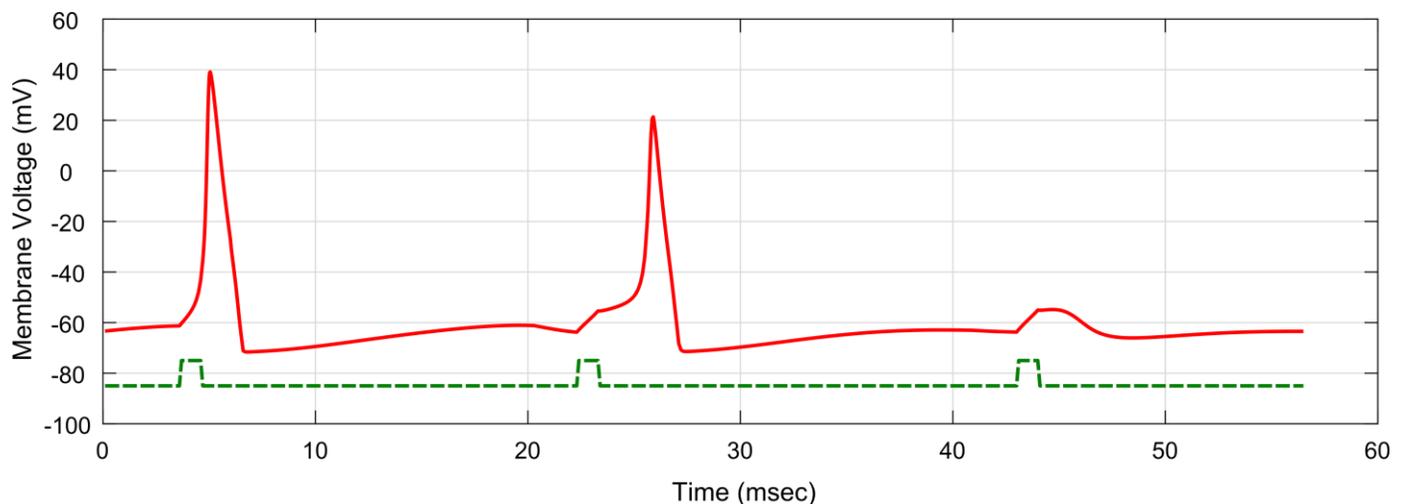

Fig 2. Response of the transmembrane potential (continuous line) of a neuronal cell segment to three consecutive equal stimuli (dashed line). Parameters in the HHSim software: conductances for Na, K, and Cl are set, respectively, to $0.0265\mu S$, $0.07\mu S$ and $0.1\mu S$. The conductance of the fast Na channel is 120 $\mu S$ for the first stimulus, 80 $\mu S$ for the second stimulus and 60 $\mu S$ for the last stimulus. Note that the AP is almost suppressed in the latter case.